\documentstyle[epsf,epsfig,twocolumn,prl,aps]{revtex}

\newcommand{\be}{\begin{equation}}
\newcommand{\ee}{\end{equation}}
\newcommand{\bea}{\begin{eqnarray}}
\newcommand{\eea}{\end{eqnarray}}

\renewcommand{\d}{{{\rm d}}}
\newcommand{\gton}{\stackrel{>}{\sim}}
\newcommand{\lton}{\mathrel{\lower.9ex
                  \hbox{$\stackrel{\displaystyle <}{\sim}$}}}

\begin{document}

\title{Explosive Decomposition in Ultrarelativistic Heavy Ion Collisions}

\author{O.\ Scavenius$^a$, A.\ Dumitru$^b$, A.D.\ Jackson$^c$}
\address{
$^a$ NORDITA, Blegdamsvej 17, DK-2100 Copenhagen {\O}, Denmark\\
$^b$ Department of Physics, Columbia University, 
538 W.\ 120th Street, New York, NY 10027, USA \\
$^c$ The Niels Bohr Institute, Blegdamsvej 17, 
DK-2100 Copenhagen {\O}, Denmark}

\maketitle   


\begin{abstract}
Recent results from Au+Au collisions at BNL-RHIC energy hint at
explosive hadron production at the QCD transition rather than soft
hydrodynamic evolution. We speculate that this is due to a rapid
variation of the effective potential for QCD close to $T_c$. 
Performing real-time lattice simulations of an effective theory
we show that the fast evolution of the potential leads to ``explosive''
spinodal decomposition rather than bubble nucleation or critical slowing
down.
\end{abstract}

\narrowtext      

It is hoped that heavy ion collision experiments at BNL-RHIC and later at 
CERN-LHC will provide evidence for the deconfined phase of QCD~\cite{expQGP}. 
Lattice gauge simulations predict a weakly first-order deconfinement 
transition at a critical temperature of $T_c \sim 
0.63\sqrt{\sigma}$, where $\sigma\sim1$~GeV/fm
is the string tension at $T=0$.  For QCD with quarks, the transition could 
remain weakly first-order or become a smooth crossover.  Even so, lattice 
data for two or three light flavors~\cite{karsch1} indicate a steep 
rise in energy density and pressure with temperature and thus 
a rapid shift of the global minimum of the effective thermodynamic potential 
for QCD within a narrow temperature interval centered at $T_c$.  To date, it 
has been difficult to identify unambiguous observables
providing compelling evidence for a phase transition in ultrarelativistic 
heavy ion collisions (URHIC).  However, rapid changes in the QCD effective 
potential near $T_c$ and/or supercooling followed by spinodal decomposition 
can have distinctive signatures.  Their experimental observation would 
probably represent the most direct information which URHIC events 
can provide regarding the QCD phase transition.  
The purpose of this Letter is to consider the likelihood of 
explosive ``spinodal'' decomposition in RHIC experiments.  

In a first-order phase transition, the thermodynamic potential for 
$T$ slightly less than $T_c$ exhibits a metastable minimum corresponding 
to the deconfined phase.  At lower temperatures, this minimum becomes a 
point of inflection, i.e.\ a ``spinodal instability''.  In a slowly 
expanding system, such as the early universe, the phase transition proceeds 
through the nucleation of bubbles of the ``true vacuum'' state via thermal 
activation for all $T < T_c$~\cite{Linde}.  In ultrarelativistic collisions 
of heavy ions or hadrons, expansion is very fast
and the metastable deconfined state may reach the inflection point 
and undergo spinodal decomposition before significant nucleation has taken 
place.  Calculations using an effective chiral model~\cite{adoveandy} 
suggest that spinodal decomposition is the favored mechanism in URHIC.  The 
condensate field in each causally connected region ``rolls down''
to its minimum with large 
attendant fluctuations in the chiral order parameter and related hadronic 
observables.

Effects similar to those of spinodal decomposition can also be expected, 
independent of the order of the phase transition, for any rapidly varying 
thermodynamic potential.  In this case, fields are unable to follow the 
rapid evolution of the equilibrium potential, and the system decays into 
any of a large number of unstable pion and sigma modes.
(By contrast, the true spinodal 
decomposition of a slowly varying potential proceeds via the ``critical'' 
growth of long-wavelength modes of the order 
parameter~\cite{Berdnikov:2000ph}.  Here, we shall use the term ``spinodal'' 
somewhat loosely to describe any rapid decomposition.)
This was first suggested in~\cite{rawil} 
in the context of chiral symmetry breaking.  It was argued that, if expansion
is sufficiently fast, the system can cool significantly while the chiral 
order parameter remains near its restored symmetry value.  Violent 
decomposition would follow as a natural consequence of such a ``quench''.  
Subsequent numerical simulations~\cite{chinum} showed that, if chiral 
symmetry breaking is considered independent of confinement, the chiral 
effective potential changes too slowly with temperature for such effects to 
be seen in URHIC.  If, however, the effective potential changes rapidly with 
temperature~\cite{adrob}, the quenched scenario and effective spinodal 
decomposition could be an interesting possibility even for moderate 
expansion rates.

RHIC ($\sqrt{s} = 130$A~GeV) has already produced a wealth of new
results~\cite{fluc}.  
``Dynamical'' fluctuations in the average transverse momentum, $p_t$, are 
larger than at the SPS ($\sqrt{s} = 17$A~GeV).  Observed 
Hanbury-Brown--Twiss (HBT) radii from pion interferometry are small, 
$\sim 5-7$~fm, and average transverse boost velocities are large, 
$\sim 0.6$c.  In particular, the fact that the HBT outward and 
sideward radii, $R_o$ and $R_s$, are roughly equal for pion pair 
transverse momenta in the range $100 < p_t < 500$~MeV may indicate 
explosive decomposition.  These phenomena all hint at a 
rapid out-of-equilibrium transition.  By contrast, if a first-order 
transition proceeds at equilibrium, a mixed phase is created, and the system 
should spend a long time at~\cite{ed} and slightly below~\cite{sven} $T_c$.  
As a result, the outward HBT radius $R_o$ should be larger than the sideward 
radius $R_s$ by as much as $50\%-100\%$~\cite{ed,sven}.

We wish to consider the possibility that the evolution of central RHIC events 
is determined by rapid changes in the thermodynamic potential.  This would 
be the case, for example, if the chiral phase transition were driven by 
the confinement transition.  The familiar rapid change of the energy density 
with temperature would then be reflected in equally rapid changes in the 
thermodynamic potential for the chiral fields.  To illustrate this point, we 
will perform real time simulations of a weakly first-order 
deconfinement transition using a recently suggested effective 
theory, valid near $T_c$, in which the free energy (or pressure) 
is dominated by a condensate of Polyakov loops $\ell$~\cite{rob}.
The potential for $\ell$ is taken as~\cite{rob}
\begin{equation} \label{ellpot}
{{{\cal V}(\ell)=\left ( -\frac{b_2}{2}
|\ell|^2-\frac{b_3}{6}(\ell^3+(\ell^*)^3)+\frac{1}{4}
(|\ell|^2)^2\right )b_4T^4}},
\end{equation}
which is invariant under global $Z(3)$ transformations.
The form of ${\cal V}(\ell)$
is dictated by usual symmetry principles~\cite{rob}.  The center 
symmetry of $SU(3)$, $Z(3)$, is broken at high $T$ where 
the global minimum of ${\cal V}(\ell)$,
$\ell_0(T)\to 1$, and restored at low $T$ where $\ell_0(T)\to 0$. 

The assumption that quarks are not important for understanding the form of the
effective potential for QCD 
is motivated by the results of lattice simulations.  Specifically, plots 
of $P/P_{\rm ideal}$ versus $T/T_c$ are remarkably insensitive to the 
number of flavors, $N_f= 0$, $2$, $3$~\cite{karsch1}.  
Thus, we neglect terms linear in $\ell$, which must exist in QCD with quarks.
Such terms would change the first-order transition (for $N_f=0$) into 
a crossover.  Nevertheless, the fact that the free energy is small 
below $T_c$ means that their numerical effects must be small.  We 
emphasize that the results (for rapid expansion)
to be described below are {\em not\/} driven 
by the order of the phase transition and would not be changed in any 
significant way by the inclusion of such linear terms.

The coefficients $b_2$, $b_3$, and $b_4$ can be chosen to reproduce 
lattice data for the pressure and energy density of pure glue theory 
for $T \ge T_c$~\cite{adrob}.  In practice, to reproduce the pressure and
energy density we may chose constant
$b_3\approx0.9$ and $b_4\approx15.0$. For $N_f=3$, we rescale $b_4$
by a factor 47.5/16, corresponding to the increase in the number of degrees
of freedom. In the spirit of mean field theory, the same constant 
values of $b_3$ and $b_4$ are to be employed for $T < T_c$.  In this domain, 
$b_2(T)\approx -0.66 \sigma^2(T)/T^4$ can be set using the finite-$T$
string tension $\sigma(T)$ from the lattice~\cite{karsch1}. 

The potential in eq.~(\ref{ellpot})
changes extremely rapidly near $T_c$, see Fig.~1 of~\cite{adrob}.   For 
temperatures even $\pm 2\%$ away from $T_c$, the potential has only 
a single minimum, which corresponds to the deconfined (confined) state.  
This rapid change in the effective potential with temperature, which is 
directly related to the rapid increase of $\sigma(T)$ below $T_c$ seen
in lattice data~\cite{karsch1}, makes ``explosive'' decomposition inevitable.
(The rapid variation of $b_2(T)$ about $T_c$ can also be understood
qualitatively by considering the above mean field theory for the second-order
confinement transition, $b_3=0$,
for the quenched theory with two colors. Then, $b_2(T)=c(T/T_c-1)$.
The rapid rise of the pressure relative to that for an ideal gas
about $T_c$~\cite{twocol}, $P(T)/P_{\rm id}(T)=b_2^2(T)$,
implies that $c$ is large.)

To complete the effective theory, we add a kinetic term for $\ell$ 
and allow coupling to a chiral field, $\phi$, describing the usual 
pions and the sigma meson~\cite{adrob}. Thus, 
\begin{equation}
{{{\cal L} \; = \; {\cal L}_\phi + 
\frac{N_c}{g^2}
|\partial_\mu \ell|^2 T^2 - {\cal V}(\ell) - \frac{h^2}{2} {\phi}^{2} \;
|\ell|^2 T^2}}\ .
\label{ec}
\end{equation}
${\cal L}_\phi$ is the standard Lagrangian for the chiral fields, see
e.g.\ \cite{adoveandy,rawil,chinum}.  The coefficient of the 
kinetic term for $\ell$ is set here by the fluctuations of $SU(3)$ Wilson 
lines in space. Perturbative corrections to that coefficient are
small~\cite{jens}.
Thus, we assume a Lorentz invariant structure~\cite{adrob}.
The coupling $h^2 \simeq 22$ between
the chiral field and $\ell$ is chosen to reproduce 
$m_\pi(T)$~\cite{gavai}.  This interaction term allows
the Polyakov loop condensate to drive the chiral phase transition.  
While the inclusion of the chiral fields 
enables us to calculate familiar observables, it is otherwise largely passive.
The behavior of $\ell$, i.e.\
confinement physics or its $Z(3)$ symmetry properties control 
the QCD transition. (In this regard, see also~\cite{igorove}.)

In our numerical simulations we solve the classical Euler-Lagrange 
equations derived from the Lagrangian~(\ref{ec}) on a $64^3$ space-like 
lattice.  We employ a Robertson-Walker metric $\d s^2= \d \tau^2 - 
a^2(\tau)\d (\tau_0\eta)^2-\d x_\perp^2$ with scale factor $a(\tau) = 
\tau/\tau_0$ and a Hubble constant $H=\d\log(a)/\d\tau=1/\tau$.
Here, $\tau$ and $\eta$ are the longitudinal proper time and the space-time
rapidity, respectively~\cite{Bj}.  We can vary the expansion rate 
by varying the initial time, $\tau_0$.  Further, 
we impose periodic boundary conditions and choose a lattice spacing 
$a_L=0.25$~fm.  We draw random initial field values at each lattice 
site from a Gaussian distribution and then coarse grain the lattice
afterwards to a correlation length of $2a_L$.  The initial 
temperature is chosen to be $T(\tau_0)=1.01 \, T_c$.  For the chiral 
field fluctuations, we assume~\cite{rawil} that $\langle \phi\rangle 
= \langle \dot{\phi} \rangle =0$, $\langle \phi^2 \rangle = v^2/4$, and 
$\langle \dot{\phi}^2 \rangle=v^2/\mbox{fm}^2$ ($v$ is the vacuum
expectation value of the sigma field, $\sigma\equiv\phi_0$). 
The $\ell$-field is initialized in the deconfined minimum of
the potential with real phase, $\langle\ell\rangle\approx0.7$. 
The magnitude of the initial fluctuations, $\langle\ell^2\rangle-
\langle\ell\rangle^2\approx0.1$ was chosen such that the
initial energy density is $\approx 4 T_c^4$~\cite{karsch1}.
The results are essentially independent of the initial
distribution of energy between fluctuations of $\ell(x)$ and $\dot{\ell}(x)$.

The presence of fluctuations modifies the potential in 
eq.~(\ref{ellpot}).  We renormalize it by expanding
${\cal V}(\ell,\phi)$ at $\tau=\tau_0$ to second order in the fluctuations
$\delta  
\ell$ and $\delta \phi$ and then subtract the resulting 
$\delta {\cal V}(\ell,\phi;\delta \ell,\delta \phi)$.  This 
provides us with the renormalized equations of motion. 

We show the results of simulations for two different 
initial times; one appropriate for ``slow'' expansion 
($\tau_0 = 2000$ fm/c) and another ($\tau_0=100$ fm/c) for ``fast'' 
expansion. 
These two cases will suffice to illustrate the radical changes which 
occur in the nature of the transition as a function of the expansion rate.
The slow expansion is, of course, many orders of magnitude faster than 
that describing the QCD transition in the early universe, while
the fast expansion is  slow on 
typical URHIC scales, and thus the indicators of non-equilibrium phenomena 
shown below should be even stronger in real URHIC.
\begin{figure}[htp]
\vspace*{-.7cm}
\centerline{\hbox{\epsfig{figure=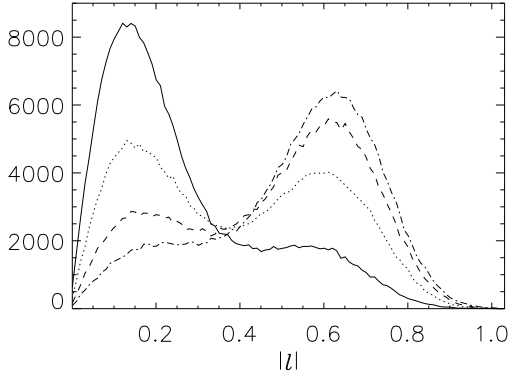,height=5.8cm}}}
\caption{$|\ell|$ field histograms obtained for the slow expansion.
The times are $a(\tau)\equiv\tau/\tau_0=1.0443$, $1.0456$, $1.0467$ and 
$1.0480$, from bottom to top in the left peak, respectively.}
\label{fig2}
\end{figure} 
The potential energy of the $\ell$-field in the
co-moving frame is assumed to red-shift 
according to $T^4/T^4_0=1/a(\tau)\equiv{\tau_0/\tau}$.
However, the spectrum of long wavelength 
fluctuations on the lattice is determined by
the dynamics of the phase transition; for example, in case of slow expansion
with nucleation and bubble growth those fluctuations on the lattice
may not cool during the phase conversion. Only hard, short wavelength
fluctuations which are integrated out and determine the coefficients in
the effective potential~(\ref{ellpot}), but are not treated explicitly in the
classical lattice simulation, cool according to the above prescribed
$T(\tau)$ law. 

Fig.~\ref{fig2} shows a sequence of histograms describing the distribution 
of $|\ell|$ for slow expansion.
The transition is weakly first order and, accordingly, the effective potential
for $\ell$ exhibits 
two local minima for $T$ in the immediate vicinity of $T_c$.  
Although $\langle \ell(x) \rangle$ is initially localized in the deconfined 
minimum at $\ell_0\sim0.7$, the fluctuations serve as a heat bath and 
trigger thermal activation over the barrier.
Fig.~\ref{fig2} shows that the confined and deconfined states coexist in 
different parts of the system. Well-defined bubbles form and, as time 
progresses, grow, coalesce, and eventually complete the 
transition to confinement.

\begin{figure}[htp]
\vspace*{-.7cm}
\centerline{\hbox{\epsfig{figure=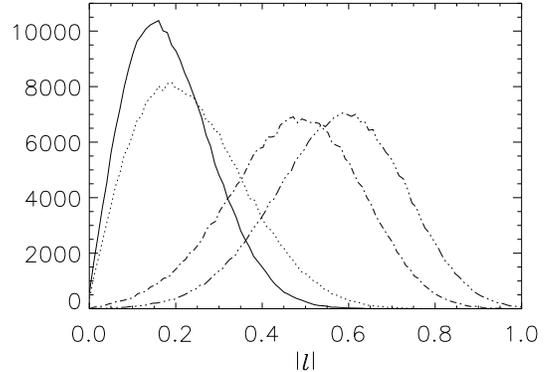,height=5.8cm}}}
\caption{$|\ell|$ field histograms obtained from the ``fast'' lattice
simulation. The times are $a(\tau)\equiv\tau/\tau_0=1.0493$, $1.0616$, 
$1.0739$ and $1.0861$ from right to left, respectively.}
\label{fig3}
\end{figure}

\begin{figure}[htp]
\vspace*{-.7cm}
\centerline{\hbox{\epsfig{figure=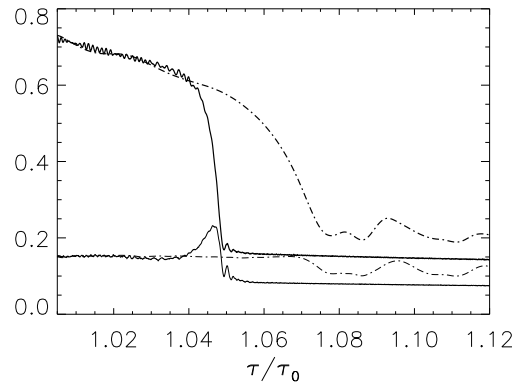,height=5.8cm}}}
\caption{The mean field, $\langle | \ell(x) | \rangle$, (thick lines) and 
its RMS fluctuations in space (thin lines).  The solid (dashed-dotted) curves
describe slow (fast) expansion.}  
\label{fig4}
\end{figure}

Fig.~\ref{fig3} shows the corresponding histograms of $|\ell|$
for the fast expansion.  Clearly, the double peak
has disappeared, and the distributions merely broaden and 
move towards the confined minimum as the system cools.  In this case, 
we conclude that the phase conversion happens through rapid
decomposition. 
For a rough estimate of the relevant time scales, note that
during a small time interval $\Delta\tau$, the hard modes cool by $|\Delta
T|/T_c \simeq \Delta\tau/4\tau_0$. From classical nucleation
theory~\cite{Linde,adoveandy} a lower bound on the time scale for nucleation is
$\Gamma^{-1/4} \gton 1/T_c$. For the model~(\ref{ellpot}) with $b_2(T)$ set by
the string tension $\sigma(T)$~\cite{karsch1}, supercooling by $\sim1\%$
suffices for the soft modes to ``roll down'' to the confined
minimum~\cite{adrob} rather than nucleation to occur. Setting
$\Delta\tau T_c\simeq1$, we expect that for $\tau_0\lton100$~fm/c phase
coexistence is not established.
Fig.~\ref{fig4} describes the time evolution of $\langle| \ell(x) |\rangle$,
and its root mean square (RMS) fluctuations.
In the slow case, we see that the mean field makes the transition from the 
deconfined minimum to the confined minimum very smoothly.
The corresponding RMS 
fluctuations increase immediately before phase conversion and
illustrate the competition between the two minima.  
(For a second-order phase transition and very slow expansion those
fluctuations would go ``critical''~\cite{Berdnikov:2000ph}).
After the transition, the mean field shows no temporal oscillations
and only small spatial fluctuations, indicating that 
the system is quite homogeneous after the transition. 
%
By contrast, fast expansion leads to strong 
temporal oscillations of the mean field and large spatial fluctuations 
following the confinement transition.  Thus, the scale of spatial homogeneity 
is smaller in this case.
 Also 
note that the rapid phase conversion and decoherence seen for fast expansion 
lead to the violent and rapid decay of the Polyakov loop condensate into 
physical particles.  This, too, suggests explosive decomposition of the 
source.  As discussed previously, strong event-by-event fluctuations of the 
pion mean-$p_t$ should emerge as a consequence~\cite{adrob}.
Also, the ``sudden'' transition could help to understand the flavor
composition in the final state~\cite{flavor}.

\begin{figure}[htp]
\vspace*{-.7cm}
\centerline{\hbox{\epsfig{figure=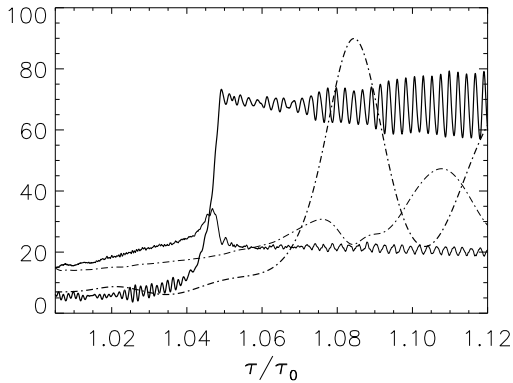,height=5.8cm}}}
\caption{The mean field $\langle \sigma(x) \rangle$ (thick lines) and RMS 
fluctuations (thin lines) in MeV.  The solid curves describe slow 
expansion.  The dashed curves describe fast expansion.}
\label{fig5}
\end{figure}

Fig.~\ref{fig5} summarizes the corresponding behavior of the chiral
$\sigma$
field.  The qualitative pattern here is the same, but the effects are 
enhanced.  This indicates that substantial energy has been flushed quickly 
from the Polyakov loop condensate into the chiral field.  This helps justify 
our expectation of the approximate equality of outward and sideward HBT 
radii: The outward radius $R_o^2$ is larger than the sideward radius $R_s^2$ 
by an amount roughly equal to the variance of the pion production time 
$\delta \tau^2 = \langle \tau^2 \rangle - \langle \tau 
\rangle^2$~\cite{ed,sven,anhen}.  For a smooth quasi-adiabatic 
transition, entropy conservation in a comoving volume element requires 
$\delta\tau^2$ to be large, while the 
non-equilibrium transition seen in our simulations is very fast.
To summarize, real-time simulations of the Polyakov loop condensate model 
have demonstrated that radically different phase transition dynamics can 
emerge in URHIC for realistically fast expansion rates when the effective 
potential changes rapidly in the vicinity of $T_c$.  The model considered 
is consistent with lattice QCD data for 
the pressure, energy density, and correlation length (or string tension) near 
$T_c$.  In real time, the 
picture that emerges is a confinement transition via spinodal decomposition 
which then triggers chiral symmetry breaking.  
 The timescales involved in URHIC seem much too short to admit
the traditional near equilibrium transition via bubble nucleation.
As expected in any 
case of phase conversion via spinodal decomposition, we observe large RMS 
spatial fluctuations in the chiral fields after the transition, i.e.\
a small spatial homogeneity length.
The near equivalence of outward and sideward HBT radii then follows as a 
natural consequence of the almost instantaneous decay of the Polyakov loop 
condensate into an ``exploding'' source of pions~\cite{anhen}.   
Large event-by-event 
fluctuations of $\langle p_t\rangle$ are another outcome of such a strong 
non-equilibrium transition.
\acknowledgements
We thank J.\ Borg, D.\ Diakonov, H.\ Heiselberg, L.\ McLerran,
I.N.\ Mishustin, 
R.D.\ Pisarski, D.\ Son, and R.\ Venugopalan for helpful criticism and
discussions.  A.D.\ 
gratefully acknowledges the support of the US-DOE through Contract 
No.\ DE-FG-02-93ER-40764.

\newpage
\begin{figure}[htp]
\vspace*{-.7cm}
\centerline{\hbox{\epsfig{figure=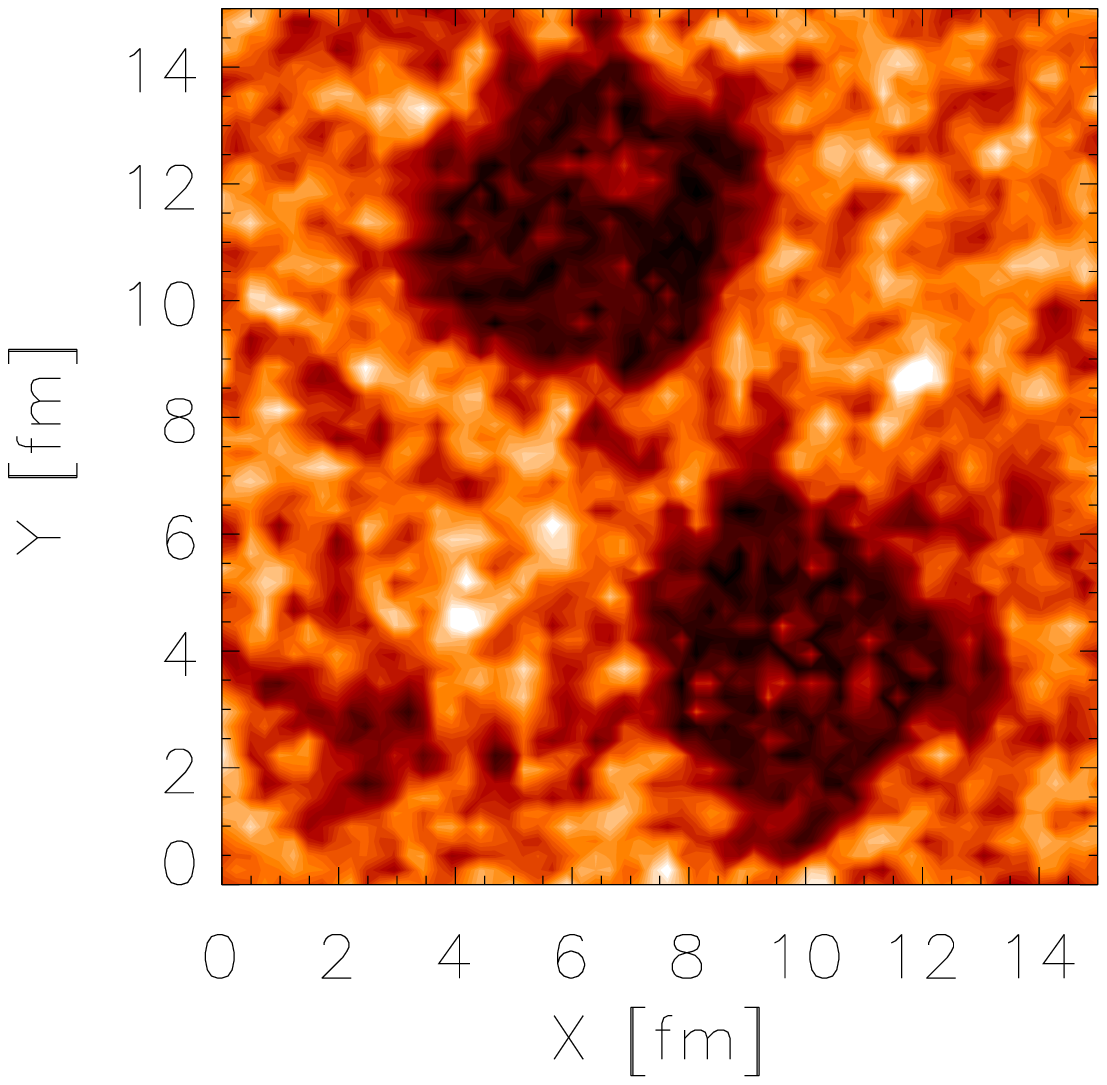,height=7cm}}}
\centerline{\hbox{\epsfig{figure=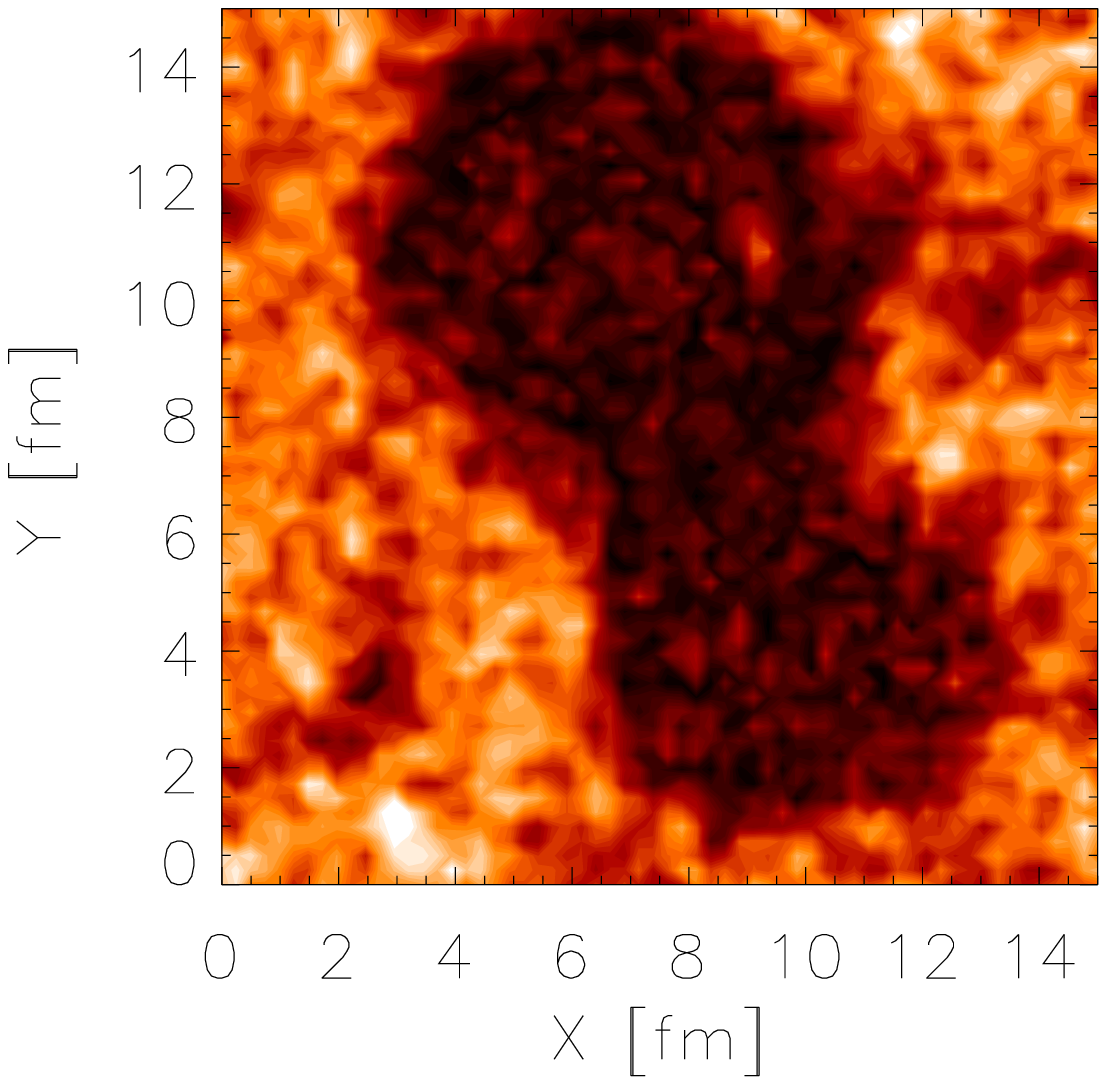,height=7cm}}}
\centerline{\hbox{\epsfig{figure=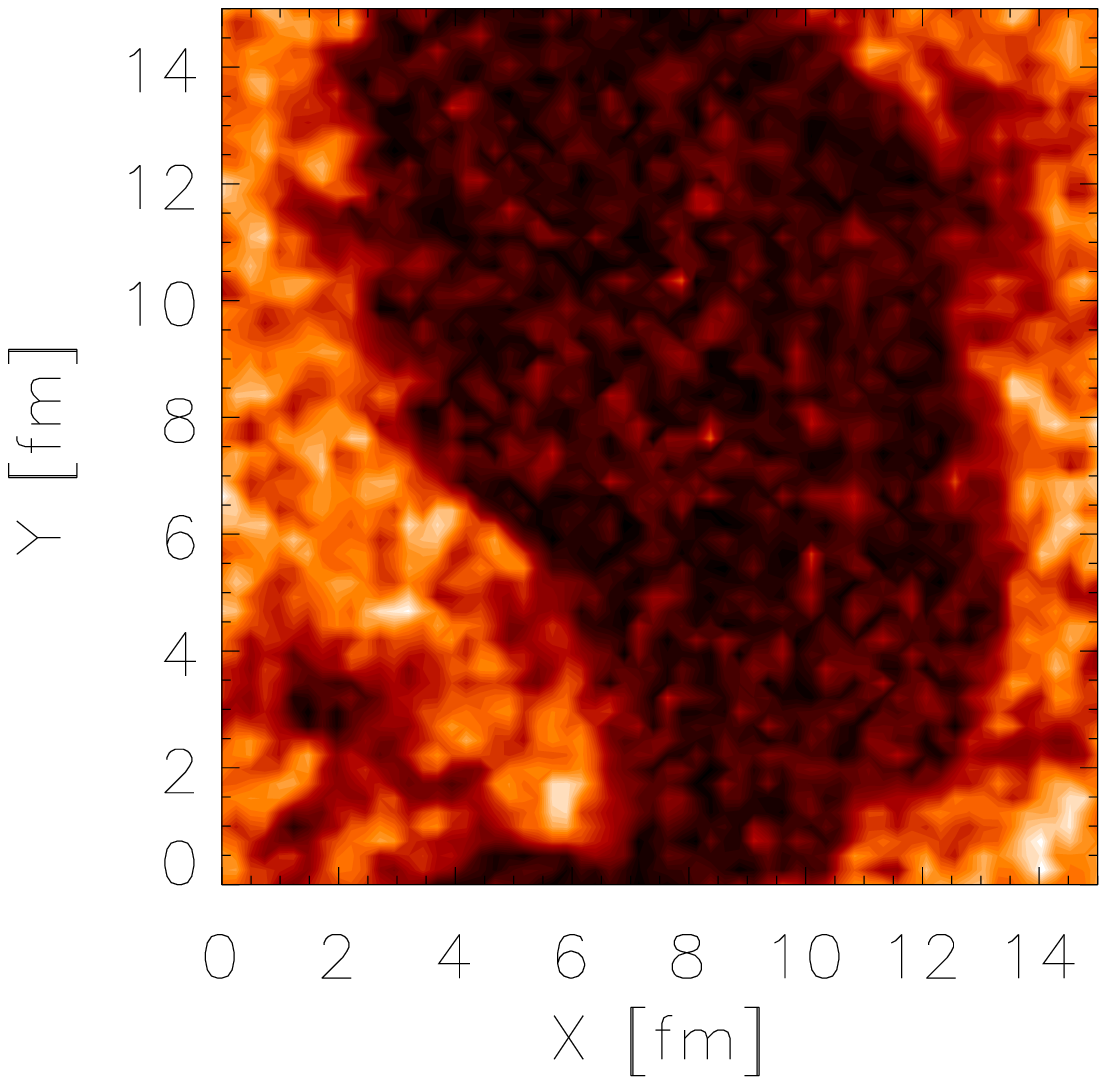,height=7cm}}}
\caption{Additional figure: $|\ell|$ contours in the transverse plane, $\eta=0$,
for the slow expansion, $\tau_0=1970$~fm.
The times are $a(\tau)\equiv \tau/\tau_0=1.045$, 1.04875, 1.05031,
from top to bottom. Dark color corresponds to small $\ell$ (confined phase)
while light color corresponds to large $\ell$ (deconfined phase).}  
\label{con1}
\end{figure} 
\begin{figure}[htp]
\vspace*{-.7cm}
\centerline{\hbox{\epsfig{figure=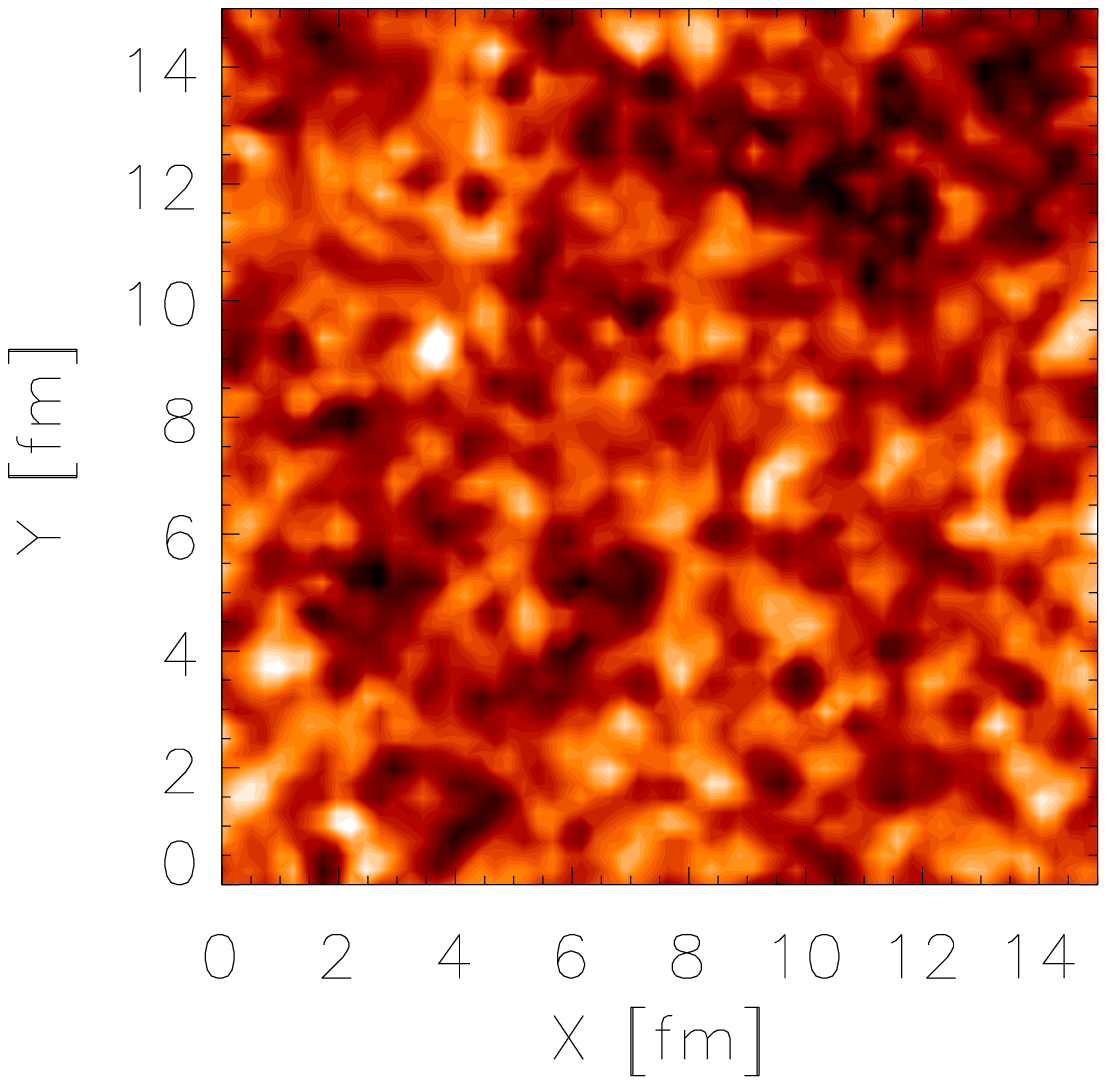,height=7cm}}}
\centerline{\hbox{\epsfig{figure=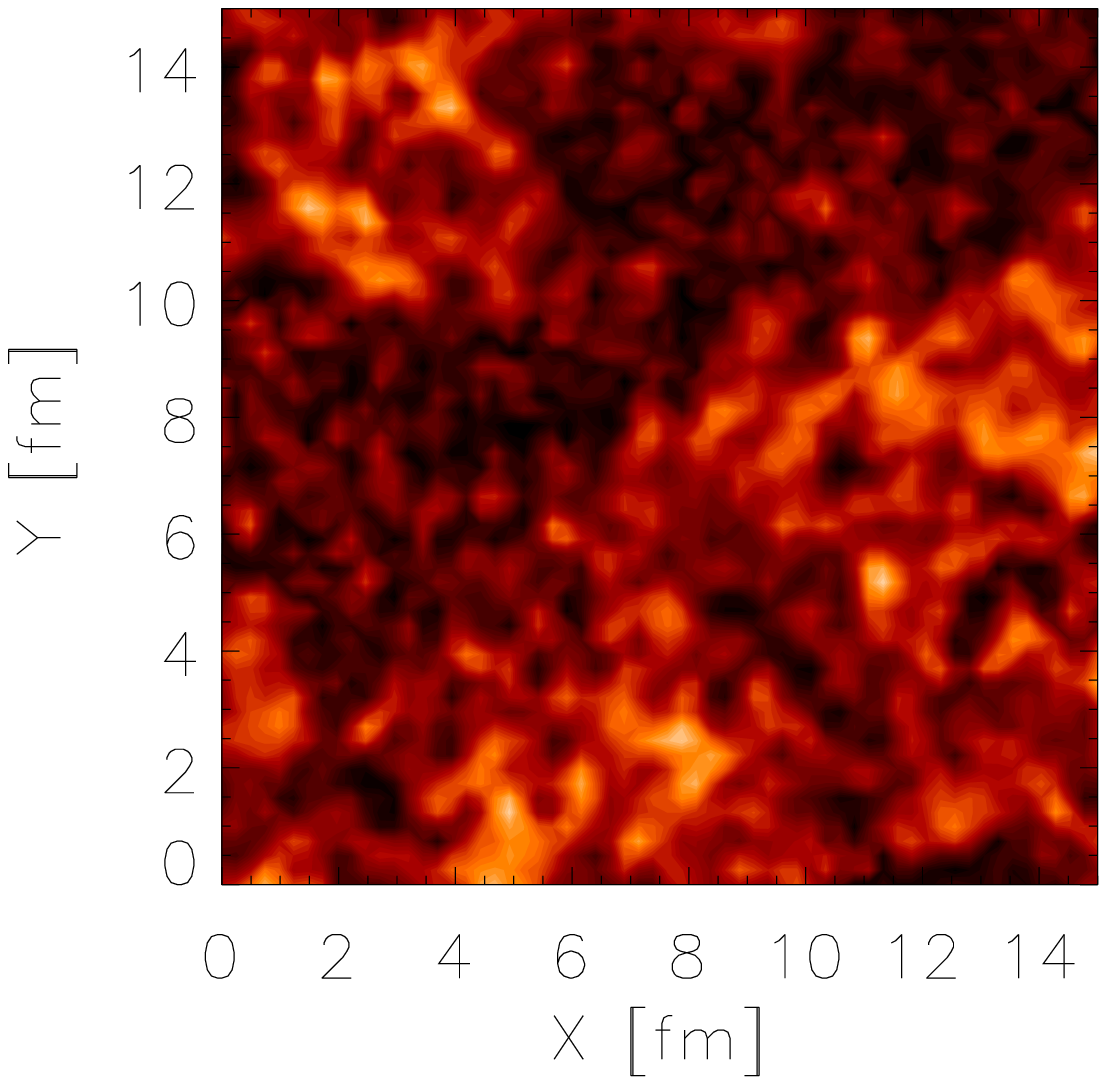,height=7cm}}}
\centerline{\hbox{\epsfig{figure=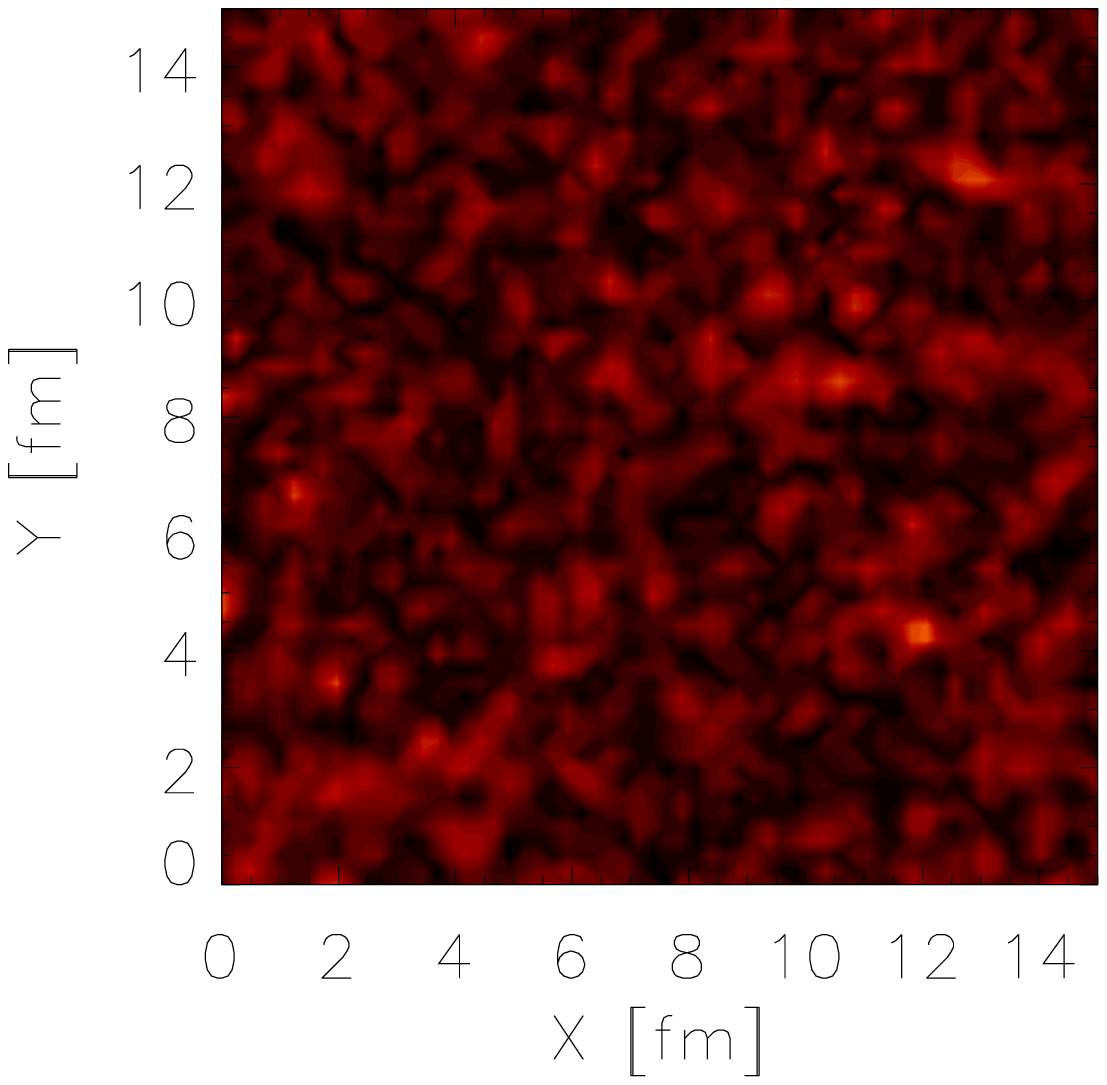,height=7cm}}}
\caption{Additional figure: $|\ell|$ contours in the transverse plane, $\eta=0$,
for the fast expansion, $\tau_0=98$~fm.
The times are $a(\tau)\equiv \tau/\tau_0=1.056$, 1.069, 1.082,
from top to bottom. Dark color corresponds to small $\ell$ (confined phase)
while light color corresponds to large $\ell$ (deconfined phase).}  
\label{con1f}
\end{figure} 
\end{document}